\documentclass[showpacs,superscriptaddress,twocolumn,prl,floatfix]{revtex4}
\usepackage{amsmath,graphicx}

\begin{document}

\title{Emission and absorption asymmetry in the quantum noise of a Josephson junction}
\author{P.-M. Billangeon}
\affiliation{Laboratoire de Physique des Solides, associ\'e au CNRS, B\^at. 510 Universit\'e Paris-Sud, 91405 Orsay Cedex, France.}
\author{F. Pierre}
\affiliation{Laboratoire de Photonique et Nanostructures, associ\'e au CNRS, Route de Nozay, 91460 Marcoussis, France}
\author{H. Bouchiat}
\affiliation{Laboratoire de Physique des Solides, associ\'e au CNRS, B\^at. 510 Universit\'e Paris-Sud, 91405 Orsay Cedex, France.}
\author{R. Deblock}
\affiliation{Laboratoire de Physique des Solides, associ\'e au CNRS, B\^at. 510 Universit\'e Paris-Sud, 91405 Orsay Cedex, France.}

\pacs{73.23.-b,05.40.Ca,42.50.Lc,72.70.+m}

\begin{abstract}
We measure current fluctuations of mesoscopic devices in the quantum regime, when the frequency is of the order of or higher than the applied voltage or temperature. Detection is designed to probe separately the absorption and emission contributions of current fluctuations, i.e. the positive and negative frequencies of the Fourier transformed nonsymmetrized noise correlator. It relies on measuring the quasiparticles photon assisted tunneling current across a superconductor-insulator-superconductor junction (the detector junction) caused by the excess current fluctuations generated by quasiparticles tunneling across a Josephson junction (the source junction). We demonstrate unambiguously that the negative and positive frequency parts of the \textit{nonsymmetrized} noise correlator are separately detected and that the excess current fluctuations of a voltage biased Josephson junction present a strong asymmetry between emission and absorption. 
\end{abstract}

\maketitle

Non equilibrium current fluctuations detection is a powerful tool to get informations not accessible by transport experiments on mesoscopic systems \cite{blanter00}. Whereas there are now a number of current noise measurements at low frequencies on different systems, there have been only few measurements in the frequency range 1-100 GHz, which is experimentally more challenging \cite{schoelkopf97,koch81,deblock03}. Such frequencies correspond to the typical energy scales and inverse propagation times involved in most mesoscopic phenomena. They correspond also to the transition to quantum noise, when $\hbar \omega \geq eV, k_B T $. Whereas book knowledge is that the measurable physical quantity in such experiments is the \textit{symmetrized} noise correlator, i.e. $C_{sym}(\tau)=\langle I(t+\tau)I(t)+I(t)I(t+\tau) \rangle$ (or alternatively the symmetrized density spectrum $S_{sym}(\omega)$ of the noise, i.e. the Fourier transform of $C_{sym}(\tau)$), several recent theoretical works have pointed out that with an appropriate detection scheme, one could measure the spectral density $S(\omega)$ \cite{Sreal} of the \textit{non-symmetrized} noise correlator $C(\tau)=\langle I(t)I(t+\tau) \rangle$ \cite{lesovik97,gavish00,aguado00}. Measuring the non-symmetrized noise means being able to distinguish between emission ($\omega < 0$, energy flows to the detector) and absorption ($\omega > 0$, energy flows from the detector) of the device under test. Experimentally it is difficult to distinguish unambiguously between symmetrized and non-symmetrized noise, partly because what is often measured is the excess noise, i.e. the difference between current fluctuations at a given bias and zero bias. In most systems studied so far at high frequencies (diffusive wire \cite{schoelkopf97}, normal tunnel junction, quantum point contact \cite{aguado00}, Josephson junction at high voltage \cite{deblock03}) the spectral density of the non-symmetrized excess noise $S^{exc}(\omega)$ is an even function of the frequency ($S^{exc}(\omega) = S^{exc}(-\omega)$) when $\hbar \omega \gg k_B T$. Consequently, the excess symmetrized noise $S_{sym}^{exc}(\omega)=S^{exc}(\omega)+S^{exc}(-\omega)$ differs only by a factor two (measured in Ref. \cite{deblock03}) from the non-symmetrized excess noise. Thus excess noise experiments in the quantum regime can usually indifferently be explained by using non-symmetrized or symmetrized noise expression. To know precisely what quantity is measured in such experiments a noise source with an asymmetric excess non-symmetrized noise is needed together with a good understanding of the detection process. In this letter we show that the current fluctuations due to quasiparticles tunneling across a Josephson junction have a clear and testable difference between absorption and emission noise. This allows us to prove that, with the detection scheme described hereafter, the emission and absorption part of non-symmetrized noise are separately measured.
\begin{figure}
	\begin{center}
		\includegraphics[width=7cm]{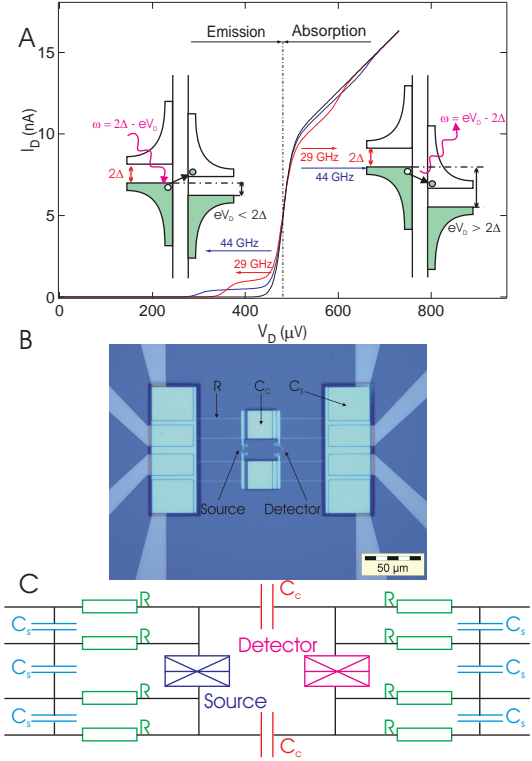}
	\end{center}
	\caption{A-Calculated $I(V)$ of the detector under monochromatic irradiation (black curve without irradiation), with frequency indicated on the figure and high enough amplitude to have a visible PAT current, and schematic pictures of the tunneling process involved in the PAT current through the detector. Left : $|V_D|< 2 \Delta/e$, only emission by the source can lead to PAT. Right : $|V_D| > 2 \Delta/e$ the detector is mainly sensitive to absorption by the source. B-Optical picture of the sample, showing the two junctions and the on-chip circuitry. C-Equivalent circuit of the sample, with $R=750$ $\Omega$, $C_C \approx 750$ fF.}
	\label{fig:figure1}
\end{figure}

The device probed in this experiment consists of two small Josephson junctions (estimated capacitance 1 fF) coupled capacitively to each other. Both junctions are made of aluminum (superconducting gap $\Delta$=240 $\mu$eV) and embedded in an on-chip environment constituted by resistances (8 Pt wires, $R$=750 $\Omega$, length=40 $\mu$m, width=750 nm, thickness=15 nm) and capacitances (estimated value $C_C \approx 750$ fF, size~: 23$\times$25 $\mu$m$^2$, insulator~: 65 nm of Al$_2$O$_3$) designed to provide a good high-frequency coupling between the two junctions (Fig. \ref{fig:figure1}). The design of the sample is similar to Ref. \cite{deblock03}, with the addition of shunt capacitances (estimated value 500 fF $< C_S <$ 750 fF) after the on-chip resistances. Those capacitances act as short-circuits at high frequencies thus isolating the on-chip circuit from the external circuit. Each junction has a SQUID geometry in order to tune the critical current with a magnetic flux. By using different SQUID areas for the two junctions the magnetic flux is adjusted to have a high critical current for one junction and a small one for the other. The junctions will hereafter be called source and detector junction. The sample is measured through filtered lines in a dilution refrigerator of base temperature 20 mK. The detection is performed by studying how the current-voltage characteristics of the detector is modified by the presence of the source and its DC polarisation \cite{deblock03}. There have been recently several proposals and/or experiments of similar detection schemes using as detector double-dots\cite{aguado00}, quantum bits \cite{schoelkopf02}, Josephson branch of a superconducting junction\cite{heikkila04,lindell04}.
\begin{figure}
	\begin{center}
		\includegraphics[clip,width=7cm]{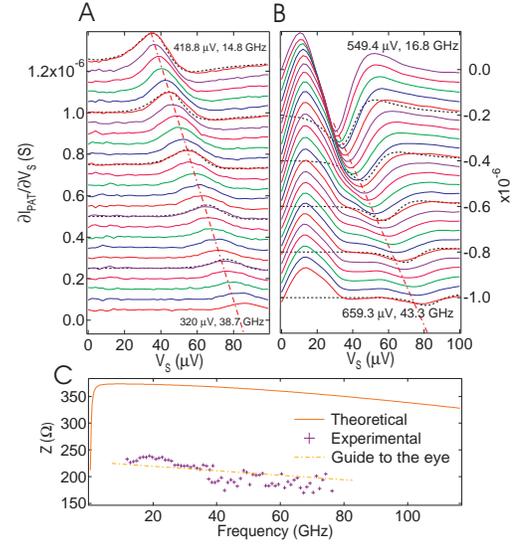}
	\end{center}
	\caption{A, B-PAT current at different detector bias (indicated on the figure) due to the AC Josephson effect ($e V_S < \Delta$). The curves are shifted vertically for clarity. A corresponds to the emission part ($V_D < 2\Delta/e$) and B to the absorption part ($V_D > 2\Delta/e$). Both exhibit a signature of the AC Josephson effect at the Josephson frequency $2 e V_S /h$ (indicated by the dot-dashed lines). The peak near $V_S=0$ in absorption is due to the variation of the source impedance on the Josephson branch. The dashed lines are calculated curves with the transimpedance as adjustable parameter. C- Transimpedance extracted from the data.}
	\label{fig:figure2}
\end{figure}

In the present work we measure the photon assisted tunneling (PAT) quasi-particle current through the detector due to the high frequency current fluctuations of the source. This current can be calculated by considering how the detector characteristic is modified by its electromagnetic environment, i.e. the source junction and the on-chip circuit \cite{ingold92,aguado00,deblock03}. Note that such a detection scheme is very general and requires only a detector with a non-linear $I(V)$ characteristic. For the particular case of a superconductor-insulator-superconductor junction the PAT current $I_{PAT}(V_D)$ of the detector, in the regime $eV_D,\hbar \omega \gg k_BT$, is~: 
\begin{eqnarray}
	&&I_{PAT}(V_D)=I_{QP}(V_D)-I_{QP,0}(V_D) \nonumber \\
	&&= \int_{0}^{+\infty} d\omega \, \left( \frac{e}{\hbar\omega} \right)^2 S_{V}(-\omega) I_{QP,0} \left( V_D+\frac{\hbar \omega}{e} \right) \nonumber \\
	&&+ \int_{0}^{eV_D} d\omega \, \left( \frac{e}{\hbar\omega} \right)^2 S_{V}(\omega) I_{QP,0} \left( V_D-\frac{\hbar \omega}{e} \right) \nonumber \\
	&&- \int_{-\infty}^{+\infty} d\omega \, \left( \frac{e}{\hbar\omega} \right)^2 S_{V}(\omega) I_{QP,0} \left( V_D \right)
	\label{Ipat}
\end{eqnarray}
with $S_{V}(\omega)$ the non-symmetrized spectral density of excess voltage fluctuations at frequency $\omega$ across the detector and $I_{QP,0}(V_D)$ the $I(V)$ characteristic of the detector when the source is not polarized. $S_{V}(\omega)$ is related to the current fluctuations of the source $S_I(\omega,V_S)$ through the transimpedance $Z(\omega)$ determined by the on-chip circuitry ($S_{V}(\omega)=|Z(\omega)|^2 S_I(\omega,V_S)$). Note that this experiment is only sensitive to the excess noise of the source, its equilibrium noise being accounted for in $I_{QP,0}(V_D)$. The different terms of Eq. \ref{Ipat} contribute only when the argument of $I_{QP,0}$ is higher than $2 \Delta/e$ (i.e. $I_{QP,0} \neq 0$). This defines two regimes of detection. When $|eV_D|<2 \Delta$ only the first term in Eq. \ref{Ipat} contributes~: we are then measuring the emission of the source (Fig. \ref{fig:figure1}A, left). When $|eV_D|>2 \Delta$, all the terms contribute but with a stronger weight for the absorption by the source (Fig. \ref{fig:figure1}A, right). Note that the measurement sensitivity is similar in emission and absorption.
\begin{figure}
	\begin{center}
		\includegraphics[clip,width=7cm]{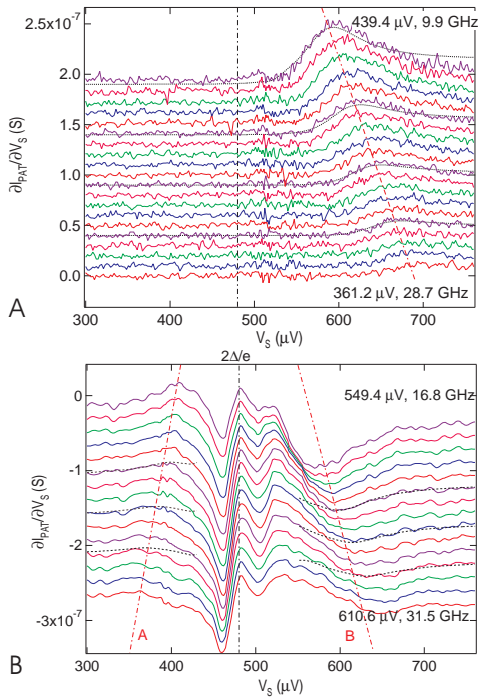}
	\end{center}
	\caption{PAT current at different bias of the detector (indicated on the figure) when the source is polarized in the vicinity of $2 \Delta$. The curves are shifted vertically for clarity. A-Emission part ($e V_D < 2 \Delta$) : there is a singularity at frequency $(e V_S-2\Delta)/h$ for $e V_S>2\Delta$ related to the emission of the source (indicated by the dot-dashed line). B-Absorption part ($e V_D > 2 \Delta$)~: there are two singularities, one at $2\Delta-e V_S$ for $e V_S < 2 \Delta$ (line A) and one at $e V_S-2\Delta$ for $e V_S > 2 \Delta$ (line B). They are related to the absorption spectrum of the source and to a change of the environment impedance (see text). The dashed lines are simulations of the expected signal taking into account the measured value of $Z(\omega)$.} 
	\label{fig:figure3}
\end{figure}

We first present data when $|e V_S| < \Delta$, with $V_S$ the bias of the source. The high frequency current across the source is then nearly monochromatic and originates from the AC Josephson effect. This current is $I(t)=I_C \sin(2 \pi \nu_J t)$ with $\nu_J = 2e V_S/h$ the Josephson frequency and $I_C=\pi \Delta/(2eR_T)$, the critical current determined by the source's normal state resistance $R_T$ \cite{tinkham96}. In the frequency domain it leads to a peak in absorption at $+\nu_J$ and one in emission at $-\nu_J$. Consequently the spectrum of the current fluctuations of the source is symmetric between emission and absorption. The PAT current through the detector is shown on Fig. \ref{fig:figure2}. To maximize sensitivity we modulate the source voltage $V_S$ and use a lock-in detection technique to measure $\partial I_{PAT}(V_D)/\partial V_S$. Whereas this quantity is not directly the derivative of the noise of the source $\partial S_I(\omega,V_S)/\partial V_S$, both look alike and display the same features, thanks to the sharp BCS density of states near the gap. Figure \ref{fig:figure2}A corresponds to the case where the detector is polarized to be sensitive only to the emission of the source ($e V_D < 2 \Delta$). For Fig. \ref{fig:figure2}B, $e V_D > 2 \Delta$ so the detector is mainly sensitive to absorption by the source. Both curves display a peak in frequency at $2e V_S/h$, which is expected from the frequency dependence of the AC Josephson effect. This similarity between emission and absorption data reflects the symmetry of current fluctuations. As done in Ref. \cite{deblock03} this signal can be used to extract the value of the transimpedance $Z(\omega)$ of the on-chip circuit versus frequency (Fig. \ref{fig:figure2}C). Despite the careful design of the on-chip circuitry, there is a difference between the measured and predicted transimpedance, which is attributed to not negligible capacitive coupling between electrodes and ground planes, or cross-talk between on-chip elements. The frequency independent peak observed in absorption close to $V_S=0$ is related to the change of the source junction impedance on the Josephson branch. This slightly changes the impedance $Z_{env}$ of the detector electromagnetic environment and thus the current of the detector \cite{ingold92}.
\begin{figure}
	\begin{center}
		\includegraphics[width=7cm]{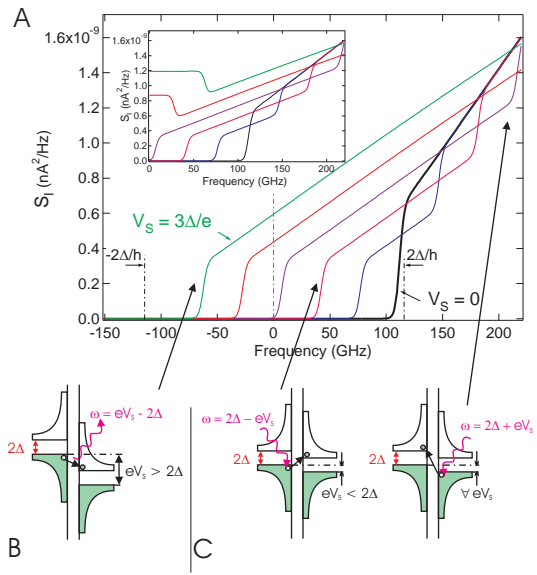}
	\end{center}
	\caption{A-Predicted non-symmetrized current fluctuations of the quasiparticle current of a Josephson junction at low temperature for different bias conditions. The highest frequency the experiment is sensitive to is $2 \Delta/h = 115$ GHz. Inset : corresponding symmetrized current fluctuations. B-Illustration of the tunneling process leading to the singularity in the emission noise ($e V_S > 2 \Delta$). C-Tunneling processes for singularities in absorption noise.}
	\label{fig:figure4}
\end{figure}

We now present data taken with the source polarized in the vicinity of the quasiparticle branch ($e V_S \approx 2 \Delta$). This regime, dominated by the shot-noise of quasiparticles, was considered to be nearly frequency independent in previous studies for bias high compared with $2 \Delta/e$ \cite{deblock03}. Our data shows a frequency dependence of this noise, both in emission (Fig. \ref{fig:figure3}A), with a singularity at frequency $(e V_S-2 \Delta)/h$ for $e V_S > 2 \Delta$, and in absorption (Fig. \ref{fig:figure3}B), in particular with a singularity at frequency $(2 \Delta -e V_S)/h$ for $e V_S < 2 \Delta$. These differences in the data reflect an asymmetry between emission and absorption. The singularities are related, \textit{via} the $I(V)$ characteristic of the source, to the superconducting density of states in the vicinity of the gap according to the calculated following expression for the non-symmetrized current fluctuations spectrum due to quasiparticle tunneling (the symmetrized spectrum was derived in Ref.~\cite{dahm69}):
\begin{eqnarray}
	S_I(\omega,V_S)=&&\frac{e}{2 \pi} \left[ \frac{I_{QP}(\hbar \omega/e + V_S)}{1-\exp{ \left( -\frac{\hbar \omega + e V_S}{k T} \right) }} \right.\nonumber \\
	&& \left. +\frac{I_{QP}(\hbar \omega/e - V_S)}{1-\exp{ \left( -\frac{\hbar \omega - e V_S}{k T} \right) }} \right]
	\label{eq1}
\end{eqnarray}
From Eq. \ref{eq1}, in the limit of low temperature, for $0 < e V_S < 2 \Delta$ there is only \textit{absorption noise} with step-like singularities (which derivative gives rise to peaks) at $\hbar \omega = 2 \Delta \pm e V_S$, whereas for $e V_S > 2 \Delta$ there are \textit{both emission and absorption} noise, with a singularity in emission at $\hbar \omega = e V_S-2 \Delta$, and a singularity in absorption at $\hbar \omega = 2 \Delta + e V_S$ (Fig. \ref{fig:figure4}). This system thus exhibits a very interesting feature in the context of noise detection~: the excess current noise has an asymmetry between emission and absorption. Note that at $V_S=0$ and $T=0$, up to frequency $2 \Delta/h$ there is no absorption noise so that excess noise and noise coincide. At higher frequencies, the excess (absorption) noise can be negative. This feature is beyond our detection range, limited to $|\hbar \omega| < 2 \Delta$.

For emission ($e V_D < 2 \Delta$), from the transimpedance and the $I(V)$ characteristic of the detector we numerically calculate the expected signal for the PAT current (dashed lines in Fig. \ref{fig:figure3}). Our calculation reproduces accurately the data, and notably the existence of a peak at $eV_S>2 \Delta$ (source emission) and the absence of peak at $eV_S<2\Delta$ (source absorption). Using Eq. \ref{eq1} one can calculate the symmetrized noise (inset of Fig. \ref{fig:figure4}). It presents the singularity in the emission noise ($e V_S > 2 \Delta$) described before but also another singularity at frequency $(2 \Delta-e V_S)/h$, when $e V_S < 2 \Delta$, which is related to the absorption side of the current fluctuations. This singularity is not present in our data for emission, proving that, when $e V_D < 2 \Delta$, only the emission part of the current fluctuations is detected.

The situation is more complicated for the absorption part ($e V_D > 2 \Delta$) because there are two contributions to the excess voltage fluctuations across the detector, leading to a modification of its characteristic $I(V_D)$ with $V_S$~: 1) the contribution of the absorption part of the excess current fluctuations spectrum of the source, given by Eq. \ref{eq1} which leads to a singularity at frequency $(2 \Delta -e V_S)/h$ for $e V_S<2 \Delta$; 2) the contribution to the excess voltage fluctuations due to the absorption current noise of the resistive part of the on-chip circuitry ($S_I(\omega)=2 \hbar \omega/(2 \pi R)$ for a resistance $R$ at $T=0$ K). Contribution 2 is not zero due to the change of the impedance $Z_{env}(\omega)$ of the detector environment with $V_S$, \textit{via} the source impedance. The source impedance is significantly modified when $e V_S \approx 2 \Delta$ (leading to the frequency independent features at $V_S\approx2 \Delta/e$ on Fig. \ref{fig:figure3}B) and presents also a high frequency dependence \cite{worsham91}, with a real part given by~: 
\begin{equation}
	\sigma(\omega,V_S)=e \frac{I_{QP}(V_S+\hbar \omega/e)-I_{QP}(V_S-\hbar \omega/e)}{2 \hbar \omega}
	\label{eq2}
\end{equation}
leading to a significant change when $\hbar \omega = \pm |2\Delta - e V_S|$. Whereas the singularity at $V_S > 2 \Delta/e$ (line B of Fig. \ref{fig:figure3}B) is only related to contribution 2, the feature at $V_S < 2 \Delta/e$ (line A) results from contribution 1 and 2, which add with different sign. This is supported by numerical calculation (dashed lines on Fig. \ref{fig:figure3}B), which reproduces qualitatively these features. The signature of the singularity in the absorption of the source junction is the reduced peak amplitude at $V_S < 2 \Delta/e$ (line A) compared with the dip at $V_S > 2 \Delta/e$ (line B). 

In conclusion we have shown in this work that by measuring the PAT current of a superconducting junction we are able to measure separately, and with comparable sensitivities, the contribution of emission and absorption to the \textit{non-symmetrized} current fluctuations of the source. For this particular detection scheme, the symmetrized current fluctuations are not relevant, because they mix emission and absorption. Whether this is a general property of any current fluctuations detection scheme in the quantum regime is an interesting issue. We also showed that the current fluctuations due to quasiparticle tunneling in a Josephson junction present a strong asymmetry between emission and absorption, with singularities in emission or absorption depending on the bias condition. 

\begin{acknowledgments} 
We acknowledge fruitful discussions with S. Gu\'eron, I. Safi and D.E. Prober. 
\end{acknowledgments}


\begin{thebibliography}{100}
\bibitem{Sreal}
Note that whereas $C(\tau)$ is not a real quantity, due to the non-commutation of the current operator at different times, $S(\omega)$ is a real quantity.
\end{thebibliography}
\end{document}